\documentclass[12pt, titlepage]{article}
\usepackage[utf8]{inputenc}
\usepackage[margin=1in]{geometry}
\usepackage{amsmath,amsfonts,bm,amssymb,bbm}
\usepackage[colorlinks,citecolor=blue,urlcolor=blue,hypertexnames=false]{hyperref}
\usepackage{booktabs}
\usepackage{setspace}
\usepackage{xcolor}
\usepackage{natbib}
\usepackage{authblk}

\newcommand\pkg[1]{\texttt{#1}}

\allowdisplaybreaks

\title{BADGERS: Bayesian Adaptive
Graph-Guided Errors-in-Variables Regression Solution for Structured
High-Dimensional Data}

\author[1]{Zeyu Hu}
\author[2]{Kewen Qu}
\author[2]{Qi Long}
\author[1]{Jun Yan}
\author[1,*]{Wenrui Li}
\affil[1]{Department of Statistics, University of Connecticut,
Storrs, CT 06269, United States}
\affil[2]{Department of Biostatistics, Epidemiology, and Informatics,
University of Pennsylvania, Philadelphia, PA 19104, United States}
\affil[*]{Correspondence:
\href{mailto:wenrui.li@uconn.edu}{Wenrui Li}}

\begin{document}
\doublespacing
\maketitle

\begin{abstract}
Knowledge-guided learning methods that incorporate biological graph information
into the analysis of structured high-dimensional data, such as genomic data, can
improve variable selection and prediction accuracy while yielding more
interpretable results. However, these methods typically assume that covariates
are measured without error. This assumption is often violated in omics studies
involving replicated measurements. To address this gap, we develop the Bayesian
Adaptive Graph-guided Errors-in-variables Regression Solution (BADGERS), a
Bayesian framework that jointly accounts for covariate measurement error and
incorporates graph information while accommodating varying numbers of replicates
across covariates. BADGERS treats the true covariates as latent variables and
integrates three key components: a Gaussian errors-in-variables model using
replicated measurements to inform the latent true covariates, a Bayesian adaptive
graphical lasso with a graph-informed prior to model their dependence structure,
and a regression model with an adaptive structured shrinkage prior to leverage
graph information and facilitate variable selection.  We develop an efficient
Markov chain Monte Carlo algorithm for posterior sampling. We demonstrate the
advantages of BADGERS over existing methods through simulation studies and
analyses of gene expression datasets for Alzheimer's disease and multiple myeloma.

\bigskip

\end{abstract}

\section{Introduction}
\label{sec:introduction}

Advances in high-throughput technologies have enabled the collection of diverse
types of omics data, including genomic and proteomic data, in biomedical
studies. Two notable examples are the Alzheimer's Disease Neuroimaging
Initiative (ADNI) study~\citep{mueller2005alzheimer} and the MicroArray Quality
Control (MAQC)-II project~\citep{shi2010maqc}. These studies provide molecular
measurements and clinical or phenotypic outcomes that can be used to develop
prediction models for disease-related outcomes, such as cognitive performance in
Alzheimer's disease (AD) and bone marrow plasma cell percentage in multiple
myeloma (MM). Many conventional predictive modeling methods do not explicitly
incorporate biological knowledge such as functional genomics and proteomics.
Extensive research has provided much information on the associations among
molecules and the underlying networks, which can be represented as graphs. For
example, biological databases such as the Kyoto Encyclopedia of Genes and
Genomes (KEGG)~\citep{kanehisa2000kegg} provide gene network information.
Knowledge-guided learning methods that incorporate biological graphs into the
analysis of such data have received increasing
attention~\citep{zhao2019knowledge, li2024knowledge}. These methods can improve
variable selection and prediction by exploiting structured relationships among
predictors.

Existing graph-guided methods generally assume that covariates are measured
without error. In practice, however, covariates are often subject to measurement 
error, particularly in omics studies involving microarray gene expression  
data, where such error may arise from various sources such as sample preparation,
labeling, and hybridization~\citep{rocke2001model,zakharkin2005sources}.
Measurement error can bias estimated covariate-response 
associations and distort inference about predictor effects, even when the
outcome is measured accurately~\citep{carroll2006measurement}. Replicate
measurements provide information about the underlying covariates and their
measurement variability. However, simply aggregating replicates may not
adequately account for measurement error, particularly when the number of
replicates differs across predictors. This issue is especially relevant in
microarray studies, where genes are measured by different numbers of probe sets.
To our best knowledge, existing work on graph-guided regression modeling has
ignored the important issue of measurement error in covariates and often uses
summary statistics such as replicate means in their modeling without explicitly
accounting for the associated noise.

As remarked above, a general framework for addressing covariate measurement
error in graph-guided regression models remains lacking. Nevertheless, there are
several areas in which errors-in-variables problem has been studied. Early work
by~\citet{durbin1954errors} employed instrumental variable techniques to reduce
coefficient estimation bias in linear regression under an additive measurement
error assumption. Other classical approaches include bias-adjusted and
functional maximum likelihood estimators for logistic
regression~\citep{stefanski1985covariate}, expectation-maximization (EM)
algorithms for generalized linear models~\citep{schafer1987covariate}, and
simulation-extrapolation estimation in parametric measurement error
models~\citep{cook1994simulation}. More recently, several methods have been
developed for high-dimensional data, including corrected loss functions and
conditional score approaches for penalized
regressions~\citep{loh2012high,sorensen2015measurement,datta2017cocolasso}, a
boosting algorithm based on measurement-error-corrected estimating
equations~\citep{brown2019meboost}, a simulation-selection-extrapolation
procedure that uses lasso in the simulation step and a group lasso for
variable selection~\citep{nghiem2019simulation}, and debiased regression
calibration for log-contrast models with compositional
covariates~\citep{zhao2024debiased}. Despite these advances, existing 
errors-in-variables methods do not incorporate external graph information to
guide regression estimation and variable selection.

Our contribution in this paper is to provide a Bayesian regression framework
that jointly accounts for covariate measurement error and leverages graph
information, while accommodating varying numbers of measurements across
covariates. The method treats the true covariates as latent and has three key
components. We employ a Gaussian errors-in-variables model, adapted
from~\citet{durbin1954errors}, that relies on replicated measurements to inform
the true underlying covariates. The latent covariate dependence structure is
modeled through the Bayesian adaptive graphical lasso approach with a 
graph-informed prior~\citep{peterson2013inferring}, which leverages network
information for estimating covariate associations. These components are
combined with a regression model relating the outcome to the latent true
covariates, equipped with an adaptive structured shrinkage prior
extending~\citet{chang2018scalable} to incorporate graph knowledge and
facilitate variable selection. This joint specification uses complementary
information from noisy covariate measurements and external knowledge graphs to
improve outcome prediction. For brevity, we refer to the proposed approach as
BADGERS, Bayesian ADaptive Graph-guided Errors-in-variables Regression
Solution. We demonstrate the advantages of our approach via simulation studies
involving varying levels of covariate measurement error and different graph
scenarios. We further illustrate its practical importance through applications
to genomic datasets for AD and MM.

The organization of this paper is as follows. In Section~\ref{sec:methodology},
we present the proposed methodology and the Markov chain Monte Carlo (MCMC) algorithm. 
Section~\ref{sec:simulation} reports the practical performance of
our method through simulation studies. In Section~\ref{sec:data_application}, 
we apply our method to two gene expression datasets. 
Finally, we conclude in Section~\ref{sec:discussion} with a discussion of future 
directions for this work. The details of the computation framework are relegated
to Appendix~\ref{sec:appendixA}.

\section{Methodology}
\label{sec:methodology}

In this section, we describe our proposed model and associated MCMC algorithm.

\subsection{Model Specification}

Suppose we have data of $n$ observations $\{y_i, \boldsymbol{w}_{i1}, \dots, 
\boldsymbol{w}_{ip} : i = 1, \dots, n\}$, where $y_i$ is the outcome 
and $\boldsymbol{w}_{ij} = (w_{ij1}, \dots, w_{ijq_{j}})^\top$ is a 
vector containing $q_j$ measurements of covariate $j$, $j = 1,\dots,p$. Let
$\boldsymbol{x}_i=(x_{i1},\dots,x_{ip})^\top$ denote the vector of latent
true covariates. Let $\mathcal{G} = (V, E)$ be the knowledge graph
for the $p$ covariates, where $V = \{1, \dots, p\}$ is the vertex set and $E
\subset \{(j,k): 1 \leq j < k \leq p\}$ is the set of undirected edges.
Denote the adjacency matrix of $\mathcal{G}$ by $\boldsymbol G=
(G_{jk})_{p\times p}$, where $G_{jk} = 1$ if there is an edge between
covariates $j$ and $k$, and $G_{jk} = 0$ otherwise.

We relate the outcome and latent true covariates through the linear model
\begin{equation}
	\boldsymbol{y} = \boldsymbol{X} \boldsymbol{\beta} + 
	\boldsymbol{\epsilon}, \quad \boldsymbol{\epsilon} 
	\sim \mathcal{N}(0, \sigma^2 \boldsymbol{I}_n),
\label{eq:outcome}
\end{equation}
where $\boldsymbol{y} = (y_1, \dots, y_n)^\top, \boldsymbol{X} = 
(\boldsymbol{x}_1, \dots, \boldsymbol{x}_n)^\top, \boldsymbol{\beta} = 
(\beta_1, \dots, \beta_p)^\top, \boldsymbol{\epsilon} = (\epsilon_1, \dots, 
\epsilon_n)^\top$, $\boldsymbol{I}_n$ denotes the $n \times n$ identity matrix, and 
$\mathcal{N}(\cdot)$ denotes the Gaussian distribution. We assume that the regression
errors are independent of the latent covariates. Adapting the errors-in-variables 
formulation of \citet{durbin1954errors}, we consider a
Gaussian measurement error model: for $i=1,\dots,n$, $j=1,\dots,p$,
$r=1,\dots,q_j$, 
\begin{equation}
	w_{ijr} = x_{ij} + \kappa_{ijr}, \quad \kappa_{ijr} \overset{\mathrm{iid}}
    {\sim} \mathcal{N}(0, \rho^2), 
\label{eq:measurement}
\end{equation}
where the measurement errors $\kappa_{ijr}$ are independent of the latent
covariates and regression errors. We assign an inverse gamma prior to the
measurement error variance $\rho^2$:
\begin{equation*}
    \rho^2 \sim \mathcal{IG}(a_{\rho},b_{\rho}),
\end{equation*}
where $\mathcal{IG}(a,b)$ denotes an inverse-gamma distribution with shape
parameter $a$ and scale parameter $b$. 
By convention, we assign the following prior to the regression error variance
$\sigma^2$:
\begin{equation*}
	\sigma^2 \sim \mathcal{IG}(a_\sigma, b_\sigma).
\end{equation*}

To model the latent covariates, we adopt the Bayesian adaptive graphical lasso
approach with a graph-informed prior proposed by \citet{peterson2013inferring}.
Following \citet{peterson2013inferring}, we specify
\begin{equation}
    \boldsymbol{x}_i \overset{\mathrm{iid}}{\sim}
    \mathcal{N}(0, \boldsymbol{\Gamma}^{-1}), \quad i=1,\dots,n,
\label{eq:latent}
\end{equation}
and assign the following prior to the precision matrix
$\boldsymbol{\Gamma}=(\gamma_{jk})_{p\times p}$,
\begin{equation}
    \pi(\boldsymbol{\Gamma} \mid \{\iota_{jk}\}_{j\leq k}) \propto \prod_{j<k}
    \left\{ \frac{\iota_{jk}}{2}\exp(-\iota_{jk} |\gamma_{jk}|)\right\}
    \prod_{j=1}^p \left\{ \frac{\iota_{jj}}{2} \exp(-\frac{\iota_{jj}}
    {2}\gamma_{jj})
    \right\} \mathbbm{1}(\boldsymbol{\Gamma} \in \mathcal{M}^+),
\end{equation} 
where $\mathbbm{1} (\cdot)$ is the indicator function and $\mathcal{M}^{+}$
denotes the set of $p\times p$ symmetric positive-definite matrices. The
shrinkage parameters $\iota_{jk}$ are treated as random variables with an
informative prior that incorporates information from the graph $\mathcal{G}$.
Specifically, we have 
\begin{equation}
    \iota_{jk} \sim \mathrm{Gamma}(a_{\iota},\, b_{\iota,jk}), 
    \quad 1 \le j \le k \le p,
\end{equation}
where $\mathrm{Gamma}(a,b)$ denotes a gamma distribution 
with shape parameter $a$ and rate parameter $b$. Let $d_{jk}$ be the length 
of the shortest path between nodes $j$ and $k$ in
$\mathcal{G}$. For $1 \le j < k \le p$, we set
\begin{equation}
b_{\iota,jk}
=
\begin{cases}
10^{-\psi} & \text{if } d_{jk}=\infty \ (\text{i.e., no path exists between nodes
} j \text{ and } k), \\[2pt]
\max\left\{10^{-\psi},\ d_{jk}^{-1}10^{-\psi+c}\right\}& \text{otherwise},
\end{cases}
\label{eq:graph-rate}
\end{equation}
where $\psi, c>0$. This specification encourages smaller
shrinkage parameters $\iota_{jk}$ for pairs of nodes that are more closely
connected in the graph $\mathcal{G}$. Smaller values of $\iota_{jk}$ impose
less shrinkage on the corresponding $\gamma_{jk}$, allowing stronger
associations between closely connected covariates. The parameter $\psi$ controls the 
overall level of shrinkage. Larger values of $\psi$ increase the prior mean of 
$\iota_{jk}$, imposing stronger shrinkage on the corresponding $\gamma_{jk}$. 
The parameter $c$ controls the strength of graph-informed shrinkage. For connected 
node pairs with $d_{jk}<10^c$, larger values of $c$ decrease the prior mean of
$\iota_{jk}$, encouraging less shrinkage on the corresponding $\gamma_{jk}$. 

We incorporate graph-guided shrinkage into the regression coefficients through
an adaptive structured prior, extending the Bayesian shrinkage approach of
\citet{chang2018scalable}. We specify 
\begin{equation}
	\beta_j \sim \mathcal{L}(0, \sigma / \lambda_j), \quad j = 1, \dots, p,
\end{equation}
where $\mathcal{L}(m,s)$ denotes a Laplace distribution with location parameter
$m$ and scale parameter $s$. The shrinkage parameters $\lambda_j$ are treated as
random variables that encode the graph knowledge $\mathcal{G}$ under an
informative network based prior. Specifically, we have
\begin{equation}
    \boldsymbol{\alpha} = (\log(\lambda_1), \dots, \log(\lambda_p))^\top 
    \sim \mathcal{N}(\mu \mathbf{1}_p, \nu \boldsymbol{\Phi}^{-1}),
\label{eq:shrinkage}
\end{equation} 
and assign a constrained Wishart prior proposed by
\citet{zhang2024incorporating} to $\boldsymbol{\Phi}$:
\begin{equation}
    \pi(\boldsymbol{\Phi}) \propto |\boldsymbol{\Phi}|^{\eta(1 + \zeta)/2}
    \exp\left\{ -\frac{\eta}{2} \mathrm{tr}\left((
    \mathbf{1}_p\mathbf{1}_p^\top + \zeta \boldsymbol{I}_p)\boldsymbol{\Phi}
    \right)\right\}   \mathbbm{1}(\boldsymbol{\Phi} \in 
    \mathcal{M}_G),
\label{eq:phi}    
\end{equation} 
where $\eta>0$, $\zeta>0$, $\mathbf{1}_p$ is the length-$p$ vector of ones,
$\boldsymbol{I}_p$ is the $p\times p$ identity matrix,  $\mathrm{tr}(\cdot)$
denotes the matrix trace, and $\mathcal{M}_G$ denotes the set of $p\times p$
symmetric positive-definite matrices whose off-diagonal $(j,k)$ entries are
nonzero if and only if $G_{jk}=1$. Our prior does not require 
$\boldsymbol{\Phi}$ to be diagonally dominant, unlike the prior proposed by
\citet{chang2018scalable}, and thus can potentially accommodate stronger
correlations. The strength of induced correlation can be controlled by adjusting
$\zeta$,  with smaller values of $\zeta$ encouraging stronger positive 
correlations between log-shrinkage parameters associated with connected nodes. 
Moreover, $\eta$ controls the concentration of the prior on $\boldsymbol{\Phi}$.
Large values of $\eta$ yield a more concentrated prior around its mode within
the graph-constrained parameter space, potentially making the posterior less 
adaptive to the data.

\subsection{Computation}
\label{sec:tuning}

We develop an efficient MCMC algorithm for sampling the parameters in our model.
To facilitate the sampling, we use a data augmentation technique that represents
the Laplace prior as a scale mixture of normal distributions. The log-shrinkage
parameter $\boldsymbol{\alpha}$ is sampled by the Metropolis–Hastings (MH)
algorithm, and all other parameters are sampled using the Gibbs sampler. The
details of the computation framework can be found in Appendix~\ref{sec:appendixA}.

We select the tuning parameters $\mu$, $\nu$, and $\psi$ by minimizing the mean 
squared prediction error (MSPE) on the validation set, defined as
\begin{equation}
    \mathrm{MSPE} = \frac{1}{|\mathcal I_{\mathrm{val}}|}\sum_{i\in\mathcal
    I_{\mathrm{val}}}
    (y_i-\widehat{y}_i)^2,\quad \widehat y_i = \frac{1}
    {|\mathcal{T}|}\sum_{t\in\mathcal{T}}
\boldsymbol{x}_i^{(t)\top}\boldsymbol{\beta}^{(t)},
\label{eq:mspe}
\end{equation}
where $\mathcal I_{\mathrm{val}}$ denotes the set of validation observations,
$\mathcal{T}$ denotes the set of retained post-burn-in MCMC iterations, and
$\boldsymbol{x}_i^{(t)}$ and $\boldsymbol{\beta}^{(t)}$ are posterior draws at
iteration $t$. All other hyperparameters are fixed in the numerical
experiments. Specifically, we set $(a_{\sigma},b_{\sigma})=(1,1)$ for
$\sigma^2$ and $(a_{\rho},b_{\rho})=(1,1)$ for $\rho^2$, which are fairly
uninformative. Following \citet{zhang2024incorporating}, we set $\eta=10$ and
$\zeta=0.2$ for $\boldsymbol{\Phi}$. We set $a_{\iota}=1$ and
$b_{\iota,jj}=10^{-2}$, $j=1,\dots,p$, following \citet{wang2012bayesian}, and
set $c=2$, as suggested by \citet{peterson2013inferring}.

We develop an open-source and user-friendly R package,
\pkg{badgers}, for implementing the proposed method. Version 0.1.0 is publicly
available on GitHub at
\url{https://github.com/zyhu888/badgers}
\citep{badgers2026}.

\section{Simulation Study}
\label{sec:simulation}

In this section, we conducted simulations to evaluate the prediction
performance of our model compared with several existing methods. We 
considered different graph scenarios and covariate noise levels.

\subsection{Simulation Setup}

The simulated outcomes were generated from the linear model~\eqref{eq:outcome},
with $\boldsymbol{x}_i \sim \mathcal{N}(0, \boldsymbol{\Sigma}_X)$ and
$\epsilon_i \sim \mathcal{N}(0, \sigma_\epsilon^2)$. The number of nonzero
coefficients in $\boldsymbol{\beta}$ was randomly chosen between 4 and 8, and
the true non-zero effect sizes were randomly generated from a uniform 
distribution with boundaries 0.5 and 2. The covariate measurements
$\boldsymbol{w}_{ij}$ were generated from the Gaussian measurement error
model~\eqref{eq:measurement} with $q_j=3$ for $i=1,\dots,n$, $j=1,\dots,p$. We
set $p=200$ and $n = 50$. The regression error variance was fixed at
$\sigma_\epsilon^2 = 2$.

Let $G_0$ denote the adjacency matrix of the true graph that governs
$\boldsymbol{X}$, where $G_{0, jk} =1$ if there is a true edge between
covariates $j$ and $k$, and $G_{0, jk} =0$ otherwise. We adapted the data
generation code of \citet{chang2018scalable} to generate $G_0$ and $\Sigma_X$,
which resembles practical gene networks encountered in applications where a
subset of important and unimportant genes lie in the same pathway and are
connected. The mechanism to generate $G_0$ and $\Sigma_X$ is described in Appendix~\ref{sec:appendixB}. We use a working graph $G$ given to fit our model, which may be
correctly specified or misspecified. We consider four scenarios involving
different working graph specifications and covariate noise levels, as follows.

\begin{enumerate}
    \item $G = G_0$; $\rho^2 = 2$.
    \item $G = G_0$; $\rho^2 = 5$.
    \item $G$ includes only a subset of the edges in $G_0$ for which the 
    corresponding partial correlations rank in the top 75\%, and randomly 
    generated edges such that the number of edges in $G$ is the same as that 
    in $G_0$; $\rho^2 = 2$.
    \item $G$ is the same as in scenario (3); $\rho^2 = 5$.
\end{enumerate}

Scenarios (1) and (2) are the cases where the true graph is completely known.
Scenarios (3) and (4)  mimic the situation where strong signals from the true
graph are known and false edges exist. Scenarios (1) and (3) involve low levels
of covariate measurement error, while scenarios (2) and (4) represent high
levels of covariate measurement error.

We generated 50 data sets, each of which contains the training set, the 
validation set, and the test set of size $n = 50$ each. The prediction
performance was assessed in terms of MSPE for the test data. The tuning
parameters $\mu$, $\nu$, and $\psi$ were selected by minimizing MSPE on the validation
set. All other hyperparameters were set to values described in
Section~\ref{sec:tuning}. We ran MCMC chains of 10,000 samples and discarded the
first 5,000 as burn-in. We used trace plots to evaluate the samples. 

\subsection{Methods to Compare}

We compared our method with the lasso (Lasso), the adaptive Lasso (ALasso), the
group Lasso (GLasso) \citep{yuan2006model,zeng2016overlapping}, the 
network-constrained regularization approach (Net) \citep{li2008network}, the EM
approach for Bayesian variable selection (EMVS) proposed by
\citet{rovckova2014emvs} and its extension incorporating structural information
(EMVSS), the EM approach for Bayesian variable selection (EMSH) proposed by
\citet{chang2018scalable} and its extension incorporating structural
information (EMSHS), the measurement error boosting approach (MEBoost) 
\citep{brown2019meboost}, the convex conditioned Lasso (CoCoLasso) 
\citep{datta2017cocolasso}, and the corrected Lasso (CorrLasso) 
\citep{loh2012high}. Among these approaches, 
GLasso treats pathways as overlapping groups, while Net, EMSHS, and EMVSS 
incorporate graph information. Since none of the competing methods can directly
accommodate multiple measurements per covariate, we used the average of
the repeated measurements for each covariate as the input. MEBoost, CoCoLasso,
and CorrLasso require the measurement error covariance matrix associated with the
averaged covariates, which we set to its true value, $(\rho^2/3)\boldsymbol{I}_p$.
The tuning parameters in these methods were chosen by minimizing the MSPE on the
validation set.

The Lasso and ALasso methods were implemented using the R Package \pkg{glmnet}
\citep{glmnetpackage}, with the initial consistent estimator for ALasso given by
ridge regression. We used the R packages \pkg{grpregOverlap}
\citep{zeng2016overlapping} and \pkg{glmgraph} \citep{chen2015glmgraph} for 
GLasso and Net, respectively. The unpublished R codes for EMVS and EMVSS were 
provided to us by the authors. The EMSH and EMSHS algorithms were implemented 
using the R code provided in the supplementary materials 
of~\citet{chang2018scalable}. MEBoost was implemented using the R code provided
in the supporting information of \citet{brown2019meboost}, and CoCoLasso and
CorrLasso were implemented using the R packages
\pkg{BDcocolasso}~\citep{escribe2021block} and
\pkg{hdme}~\citep{sorensen2019hdme}, respectively.

\subsection{Results}

The MSPEs for the test data are reported in Table~\ref{table:1}. BADGERS
outperforms all competing methods across all settings, including different
levels of covariate measurement error and working graph specifications.

We first evaluate the effect of covariate measurement error on prediction
performance. Under low measurement error (scenarios (1) and (3)), methods
incorporating graph information (i.e., EMVSS, EMSHS, and BADGERS) achieve lower
mean MSPEs than methods that do not incorporate graph information. Under high
measurement error (scenarios (2) and (4)), BADGERS has superior prediction
performance compared to all competing methods. In particular, the improvement 
of BADGERS over EMVSS and EMSHS is more pronounced under high measurement
error. These results suggest that BADGERS can improve prediction accuracy by
accounting for covariate measurement error.

We next examine the sensitivity of prediction performance to working graph
misspecification. Compared with the correctly specified graph settings
(scenarios (1) and (2)), the mean MSPEs of BADGERS increase modestly under
graph misspecification (scenarios (3) and (4)). Notably, under high covariate
measurement error, BADGERS with a misspecified working graph (scenario (4))
achieves a lower mean MSPE than all competing methods with the correctly
specified graph (scenario (2)). These findings suggest that BADGERS is
relatively robust to working graph misspecification.

\begin{table}[tbp]
\doublespacing
\centering
\caption{Mean MSPEs on the test sets, with standard deviations in parentheses.}
\label{table:1}
\begin{tabular}{lcccc}
\toprule
Method & Scenario (1) & Scenario (2) & Scenario (3) & Scenario (4) \\
\midrule
Lasso & 12.066 (4.620) & 23.921 (10.955) & 12.066 (4.620) & 23.921 (10.955) \\
ALasso & 11.192 (4.367) & 23.439 (10.837) & 11.192 (4.367) & 23.439 (10.837) \\
GLasso & 11.054 (4.393) & 21.085 (8.586) & 11.160 (4.395) & 21.180 (8.729) \\
Net & 10.481 (4.073) & 20.977 (9.718) & 10.914 (4.006) & 21.505 (9.763) \\
EMVS & 10.513 (4.281) & 23.697 (10.842) & 10.513 (4.281) & 23.697 (10.842) \\
EMVSS & 10.168 (4.062) & 21.694 (9.576) & 10.045 (4.103) & 21.844 (10.226) \\
EMSH & 11.720 (4.878) & 24.753 (12.907) & 11.720 (4.878) & 24.753 (12.907) \\
EMSHS & 9.281 (3.631) & 20.371 (12.183) & 10.240 (3.983) & 21.706 (11.857) \\
MEBoost & 12.424 (5.476) & 24.746 (11.657) & 12.424 (5.476) & 24.746 (11.657) \\
CoCoLasso & 11.894 (4.828) & 24.410 (12.909) & 11.894 (4.828) & 24.410 (12.909) \\
CorrLasso & 17.461 (7.537) & 26.910 (12.153) & 17.461 (7.537) & 26.910 (12.153) \\
BADGERS & 9.016 (3.457) & 17.263 (7.366) & 9.713 (3.860) & 19.051 (8.819) \\
\bottomrule
\end{tabular}
\end{table}

\section{Data Application}
\label{sec:data_application}

We applied the proposed method to analyses of two gene expression datasets, one
from the ADNI study~\citep{mueller2005alzheimer} and the other from an MM study in
the MAQC-II project~\citep{shi2010maqc}.

\subsection{Application to an ADNI Study}
\label{sec:adni}

The ADNI is a large-scale, multisite longitudinal study that tracks the
progression of AD from normal aging through mild cognitive impairment (MCI) to
dementia \citep{mueller2005alzheimer}. The ADNI study aims to validate
diagnostic and prognostic biomarkers for AD progression. In our application, we
investigated the association of patients' gene expression levels with their
Montreal Cognitive Assessment (MoCA) scores. The MoCA is a cognitive screening test
designed to detect MCI and early dementia by assessing multiple cognitive
domains, including memory, executive function, attention, concentration,
language, and orientation. MoCA scores range from 0 to 30, with lower scores
indicating greater cognitive impairment.

To prepare the ADNI data for our analysis, we applied a Box-Cox transformation
to the outcome variable MoCA score for normalization, and then standardized
both the outcome and covariates. As a screening step, we conducted univariate
regression of each gene in relation to the MoCA score and selected genes with a
false discovery rate (FDR) $<0.01$ for subsequent analyses. There are 264
selected genes. Among these genes, 30\%, 26\%, 21\%, and 23\% were measured
using one, two, three, and more than three probes, respectively. The gene
network $G$ was retrieved from the KEGG database \citep{kanehisa2000kegg}.
There are 722 subjects in the analysis. We randomly divided them into training,
validation, and test sets of sizes 240, 240, and 242, respectively. We fitted
our method and the competing methods considered in the simulation study and
recorded the MSPE for the test set. The tuning parameters for the competing
methods were selected by minimizing the MSPE on the validation set.  For 
MEBoost, CoCoLasso, and CorrLasso, motivated
by~\citet{carroll2006measurement}, we estimated the measurement error 
covariance matrix for the averaged covariates within each training set as 
$\operatorname{diag}(\widehat{\rho}^2/q_1,\ldots,\widehat{\rho}^2/q_p)$, 
where $\widehat{\rho}^2$ is the pooled within-subject, within-covariate
replicate variance. For our proposed approach, we used the prior
specifications and tuning procedure described in Section~\ref{sec:tuning}. We
repeated the analysis 50 times. 

\begin{table}[tbp]
\doublespacing
\centering
\caption{Mean MSPEs for the ADNI data set, with standard deviations in parentheses.}
\label{table:2}
\begin{tabular}{l@{\hspace{1.5cm}}c} 
\toprule
Method & MSPE \\
\midrule
Lasso & 0.953 (0.075) \\
ALasso & 0.966 (0.078) \\
GLasso & 0.966 (0.077) \\
Net & 0.955 (0.076) \\
EMVS & 0.999 (0.077) \\
EMVSS & 1.005 (0.092) \\
EMSH & 0.966 (0.084) \\
EMSHS & 0.967 (0.083) \\
MEBoost & 0.942 (0.078) \\
CoCoLasso & 0.982 (0.083) \\
CorrLasso & 0.987 (0.077) \\
BADGERS & 0.936 (0.069) \\
\bottomrule
\end{tabular}
\end{table}

Table~\ref{table:2} shows the MSPEs for the test set. BADGERS yields the best
prediction performance, which demonstrates the advantage of our method. In
addition, we used 80\% credible intervals to select genes and performed the
pathway enrichment analysis via the ToppGene Suite \citep{chen2009toppgene}.
The selected genes, such as CD177 and OAZ3, have been shown to be associated
with AD \citep{le2017polymorphonuclear, iturria2022unified}. Moreover, several
enriched pathways have been shown to be associated with AD, including serine
metabolism~\citep{ledouce2020impairment}, fibrin clot
formation~\citep{cortescanteli2010fibrinogen}, and polyamine
metabolism~\citep{mahajan2020dysregulation}. These findings suggest that BADGERS
yields biologically meaningful results.

\subsection{Application to an MM Study}

The MAQC-II project focused on development and evaluation of microarray-based
predictive models~\citep{shi2010maqc}. The project included an MM dataset
contributed by the Myeloma Institute for Research and Therapy at the University
of Arkansas for Medical Sciences, which contains gene expression profiles of
CD138-purified bone marrow plasma cells from newly diagnosed patients with MM. In
our application, we investigated the association of patients' gene expression
levels with their bone marrow aspirate plasma cell percentage (ASPC). ASPC is
calculated as the percentage of plasma cells in a bone marrow aspirate and
reflects the extent of plasma-cell infiltration in the bone marrow.

To prepare the MM dataset for analysis, we standardized the outcome and
covariates. As a screening step, we conducted univariate regression of each
gene in relation to the ASPC and selected genes with FDR $<0.001$ for subsequent 
analyses. There are 222 selected genes. Among these genes, 43\%, 27\%, 13\%, 
and 17\% were measured using one, two, three, and more than three probes, 
respectively. The gene network $G$ was obtained from the
KEGG database~\citep{kanehisa2000kegg}. There are 531 patients in the analysis.
We randomly divided them into training, validation, and test sets of equal size.
We fitted our method and the competing methods
considered in the simulation study and recorded the MSPE for the test set. 
For MEBoost, CoCoLasso, and CorrLasso, the measurement error covariance 
matrix was estimated using the same approach as in Section~\ref{sec:adni}. 
The tuning parameters for the competing methods were selected by minimizing 
the MSPE on the validation set. For our proposed approach, we used the 
prior specifications and tuning procedure described in 
Section~\ref{sec:tuning}. We repeated the analysis 50 times.

\begin{table}[tbp]
\doublespacing
\centering
\caption{Mean MSPEs for the MM data set, with standard deviations in parentheses.}
\label{table:3}
\begin{tabular}{l@{\hspace{1.5cm}}c} 
\toprule
Method & MSPE \\
\midrule
Lasso & 0.767 (0.064) \\
ALasso & 0.782 (0.065) \\
GLasso & 0.843 (0.061) \\
Net & 0.764 (0.063) \\
EMVS & 0.978 (0.089) \\
EMVSS & 0.820 (0.077) \\
EMSH & 0.786 (0.067) \\
EMSHS & 0.782 (0.067) \\
MEBoost & 0.757 (0.062) \\
CoCoLasso & 0.814 (0.069) \\
CorrLasso & 0.829 (0.081) \\
BADGERS & 0.749 (0.063) \\
\bottomrule
\end{tabular}
\end{table}

Table~\ref{table:3} shows the MSPEs for the test set. BADGERS achieves the lowest
MSPE, demonstrating its prediction advantage. Using the same 80\% credible
interval criterion and pathway enrichment procedure described in
Section~\ref{sec:adni}, we selected several genes and identified enriched
pathways relevant to MM. PTHLH encodes parathyroid hormone-related protein, which
promotes myeloma plasma cell survival and pro-osteoclast
activity~\citep{cafforio2014pthrp}. Several selected immunoglobulin variable
region genes, including IGKV2D-28, IGLV3-25, and IGLV3-10, have been observed in
immunoglobulin repertoires from patients with
MM~\citep{hadzidimitriou2006immunoglobulin}. In addition, several enriched
pathways, including B-cell receptor signaling and the complement cascade, have
been implicated in MM~\citep{yan2019identification,sanchez2016daratumumab}. These
findings show that BADGERS yields biologically meaningful results.    

\section{Discussion}
\label{sec:discussion}

We have developed a Bayesian framework for predicting an outcome from
structured high-dimensional data with covariate measurement error. Our approach
jointly leverages information from replicate covariate measurements and
biological knowledge graph extracted from database to improve prediction
accuracy. Simulation studies demonstrate the superiority of our proposed method
over several existing methods. Analyses of two real data sets further
illustrate the prediction performance of our method and its ability to
identify biologically relevant genes and pathways.

Our work sets the stage for several extensions. First, extending our approach
to generalized linear models and other modeling frameworks would broaden its
applicability. Second, it would be useful to accommodate more general
distributions of measurement error and covariates. Third, jointly accounting
for noise in the graph information and covariates could further improve
prediction performance when nontrivial noise is present in the graph. Finally,
developing alternative algorithms, such as variational EM, could reduce the
computational cost.

\appendix
\section{MCMC Algorithm}
\label{sec:appendixA}

In this section, we describe the MCMC algorithm for sampling the parameters in
our model. We first introduce a data augmentation technique to facilitate the
sampling. We then present the joint posterior distribution and the conditional
posterior distributions of the model parameters. The log-shrinkage parameter
$\boldsymbol{\alpha}$ is sampled by the Metropolis–Hastings (MH) algorithm, and
all other parameters are sampled using the Gibbs sampler.

\subsection*{A.1 Data Augmentation}
\label{sec:data_aug}

To facilitate posterior sampling in our MCMC algorithm, we employ a data
augmentation technique that represents the Laplace prior as a scale mixture of
normal distributions.

The Laplace prior for $\beta_j$ can be represented as a scale mixture of 
normals with an exponential mixing density \citep{andrews1974scale}
\begin{align*}
     \frac{\lambda_{j}}{2\sigma} \exp\left(-\frac{\lambda_{j}\lvert 
     \beta_{j}\rvert}{\sigma} \right)
     = \int_0^\infty \frac{1}{\sqrt{2\pi s\sigma^2}}  
     \exp\left(- \frac{\beta_{j}^2 }{2s\sigma^2}\right) \frac{\lambda_{j}^2 }
     {2} \exp\left(-\frac{\lambda_{j}^2s}{2}\right) ds.
\end{align*}
This suggests the following representation of the prior for $\beta_{j}$:
\begin{align*}
     \beta_{j}\mid\tau_{j}^2,\sigma^2 \sim \mathcal{N}
     (0, \sigma^2 \tau_j^2), \quad
     \tau_j^2\mid\lambda_{j} \sim \mathrm{Exp}(\lambda_j^2/2), 
     \quad j = 1,\dots,p,
\end{align*}
where $\mathrm{Exp}(a)$ denotes the exponential distribution with rate
parameter $a$. 

Similarly, the Laplace priors on the off-diagonal elements of
$\boldsymbol{\Gamma}$ can be represented as 
\begin{align*}
    \gamma_{jk} \mid \upsilon_{jk}^2 \sim \mathcal{N}(0, 
    \upsilon_{jk}^2), \quad \upsilon_{jk}^2 \mid \iota_{jk} \sim \mathrm{Exp}
    (\iota_{jk}^2/2), \quad 1\leq  j < k \leq p.
\end{align*}

\subsection*{A.2 Conditional Posterior Distributions}
\label{sec:post}

Let $n$ denote the number of subjects in the training set, and let 
$\boldsymbol{\Theta}$ denote the collection of all parameters in the model.
The joint posterior density given training data is 
\begin{align*}
\begin{split}
    \pi(\boldsymbol{\Theta} \mid \boldsymbol{y}, \boldsymbol{W}) \propto{}&
    \pi(\boldsymbol{y}  \mid \boldsymbol{X}, \boldsymbol{\beta}, \sigma^2)
    \pi(\boldsymbol{\beta} \mid \boldsymbol{\tau}^2, \sigma^2)  
    \pi(\boldsymbol{\tau}^2\mid \boldsymbol{\alpha}) \pi(\boldsymbol{\alpha} 
    \mid \boldsymbol{\Phi}) \pi(\boldsymbol{\Phi} ) 
    \pi(\sigma^2) \pi(\rho^2)\\
    &\times  \left[  \prod_{i=1}^n \prod_{j=1}^p \prod_{r=1}^{q_j} \pi(w_{ijr}
    \mid x_{ij}, \rho^2) 
    \right] \left[ \prod_{i=1}^n \pi(\boldsymbol{x}_i \mid \boldsymbol{\Gamma}) 
    \right] \pi(\boldsymbol{\Gamma} \mid \boldsymbol{\Upsilon}, 
    \{\iota_{jj}\}_{j=1}^p) \\
    &\times  \left[ \prod_{1 \le j < k \le p} \pi(\upsilon_{jk}^2 \mid
    \iota_{jk}) \right] 
    \left[ \prod_{1 \le j \le k \le p} \pi(\iota_{jk} )
    \right]
\end{split}
\end{align*}
where $\boldsymbol{W} = \{\boldsymbol{w}_{ij}:i=1,\dots,n,\ j=1,\dots,p \}$,
$\boldsymbol{\tau}^2 = (\tau_1^2, \ldots, \tau_p^2)^\top$, and
$\boldsymbol{\Upsilon} = (\upsilon_{jk}^2)_{p\times p}$ is a symmetric matrix
with a zero diagonal.

Let $\boldsymbol{y}^*$ and $\boldsymbol{W}^*$ denote the outcomes and observed
replicate covariate measurements, respectively, in the validation or test set.
The posterior predictive distribution of $\boldsymbol{y}^*$ given
$\boldsymbol{W}^*$ and the training data is
\begin{align*}
\begin{split}
    &\pi(\boldsymbol{y}^* \mid \boldsymbol{W}^*, \boldsymbol{y},
    \boldsymbol{W}) \\
    ={}& \int\int
    \pi(\boldsymbol{y}^* \mid \boldsymbol{X}^*, \boldsymbol{\Theta})
    \pi(\boldsymbol{X}^* \mid \boldsymbol{W}^*, \boldsymbol{\Theta})
    \pi(\boldsymbol{\Theta} \mid \boldsymbol{y}, \boldsymbol{W})
    \,d\boldsymbol{X}^*\,d\boldsymbol{\Theta},
\end{split}
\end{align*}
where $\boldsymbol{X}^*$ denotes the latent covariates in the validation or
test set.

Next, we show the full conditional posterior distributions of the model
parameters and $\boldsymbol{X}^*$. 

\textit{(1) The conditional distribution of $\boldsymbol{\beta}$}. We use 
Gibbs sampling to sample $\boldsymbol{\beta}$ from its conditional 
distribution, which is 
\begin{align*}
\begin{split}
    \boldsymbol{\beta} \mid \boldsymbol{X}, \boldsymbol{y}, \sigma^2, 
    \boldsymbol{\tau}^2 \sim \mathcal{N}\left(\boldsymbol{A}
    \boldsymbol{X}^{\top}\boldsymbol{y}/\sigma^2, \boldsymbol{A} \right),
\end{split}
\end{align*}
where $\boldsymbol{D}_{\boldsymbol{\tau}} = \mathrm{diag}(\tau_1^2,\ldots,
\tau_p^2)$ and $\boldsymbol{A} = \sigma^2\left(\boldsymbol{X}^{\top}
\boldsymbol{X} + \boldsymbol{D}_{\boldsymbol{\tau}}^{-1}\right)^{-1}$.

\textit{(2) The conditional distributions of $\tau_j^2$}. We use Gibbs 
sampling to sample $\tau_j^2$ from its conditional distribution, that is 
\begin{align*}
    \frac{1}{\tau_j^2} \mid \beta_j, \lambda_{j}, \sigma^2 \sim \mathcal{IN}
    \left(\frac{\lambda_j\sigma}{\vert \beta_j \vert }, \lambda_j^2 \right),
    \ j= 1,\dots, p,
\end{align*}
where $\mathcal{IN}(\mu,\lambda)$ refers to the inverse-Gaussian distribution with parameters $\mu$ and $\lambda$ in the parameterization of the inverse Gaussian density given by
\begin{align*}
f(x)=\sqrt{\frac{\lambda}{2\pi}} x^{-3/2}\exp\left\{ -\frac{\lambda(x-\mu)^2}{2\mu^2x} \right\},\ x> 0  . 
\end{align*}

\textit{(3) The conditional distribution of $\sigma^2$}. We use Gibbs 
sampling to sample $\sigma^2$ from its conditional distribution, which is 
\begin{align*}
    \sigma^2 \mid \boldsymbol{y}, \boldsymbol{X}, \boldsymbol{\beta}, 
    \boldsymbol{\tau}^2 \sim \mathcal{IG}\left(a_\sigma + \frac{n+p}{2},\; 
    b_\sigma + \frac{(\boldsymbol{y}-\boldsymbol{X}
    \boldsymbol{\beta})^\top(\boldsymbol{y}-\boldsymbol{X}\boldsymbol{\beta}) 
    + \boldsymbol{\beta}^\top 
    \boldsymbol{D}_{\boldsymbol{\tau}}^{-1}\boldsymbol{\beta}}{2}\right).
\end{align*}

\textit{(4) The conditional distribution of $\boldsymbol{\alpha}$}. We use 
MH algorithm to sample $\boldsymbol{\alpha}$ from its conditional 
distribution, that is 
\begin{align*}
    \pi(\boldsymbol{\alpha}
    \mid \boldsymbol{\tau}^2, 
    \boldsymbol{\Phi})
    \propto
    \exp\left\{
        \sum_{j=1}^{p}
        \left[
            2\alpha_j
            - \frac{\exp(2\alpha_j)\tau_j^2}{2}
        \right]
        - \frac{1}{2\nu}
        (\boldsymbol{\alpha}-\mu\mathbf{1}_p)^\top
        \boldsymbol{\Phi}
        (\boldsymbol{\alpha}-\mu\mathbf{1}_p)
    \right\}.
\end{align*}

\textit{(5) The conditional distribution of $\boldsymbol{\Phi}$}. We use a 
block-wise Gibbs sampler to sample $\boldsymbol{\Phi}$ column by column. Let
$\phi_{jj}$ be the $j$th diagonal element of  $\boldsymbol{\Phi}$,
$\boldsymbol{\phi}_{-j,j}$ be the vector of off-diagonal elements in its $j$th
column, and $\boldsymbol{\Phi}_{-j,-j}$ be the submatrix of $\boldsymbol{\Phi}$
excluding the $j$th row and column, so that we have, after implicit row and
column permutations,
\begin{align*}
    \boldsymbol{\Phi} = \begin{bmatrix} \boldsymbol{\Phi}_{-j,-j} & 
    \boldsymbol{\phi}_{-j,j} \\ \boldsymbol{\phi}_{-j,j}^\top & \phi_{jj} 
    \end{bmatrix}.
\end{align*}
Similarly, after the same implicit permutations, we have 
\begin{align*}
    \boldsymbol{A}_{\Phi} = \begin{bmatrix} \boldsymbol{A}_{\Phi,-j,-j} & 
    \boldsymbol{a}_{\Phi,-j,j} \\ \boldsymbol{a}_{\Phi,-j,j}^\top & 
    a_{\Phi,jj} \end{bmatrix},
\end{align*}
where $\boldsymbol{A}_{\Phi} = \eta\left(\mathbf{1}_{p}\mathbf{1}_{p}^\top + 
\zeta \boldsymbol{I}_{p}\right) + \frac{1}{\nu}(\boldsymbol{\alpha}-
\mu\mathbf{1}_p)(\boldsymbol{\alpha}-\mu\mathbf{1}_p)^\top$, and
$\boldsymbol{A}_{\Phi,-j,-j}$, $\boldsymbol{a}_{\Phi,-j,j}$ and $a_{\Phi,jj}$ 
denote the corresponding submatrix, vector, and diagonal element of
$\boldsymbol{A}_{\Phi}$, respectively. Let $\boldsymbol{\phi}_{-j,j}^1$ 
($\boldsymbol{\phi}_{-j,j}^0$) be the subvector of nonzero (zero) entries in 
$\boldsymbol{\phi}_{-j,j}$, and 
$\boldsymbol{a}_{\Phi,-j,j}^1$ ($\boldsymbol{a}_{\Phi,-j,j}^0$) be the 
subvector of $\boldsymbol{a}_{\Phi,-j,j}$ corresponding to 
$\boldsymbol{\phi}_{-j,j}^1$ ($\boldsymbol{\phi}_{-j,j}^0$), and 
$\widetilde{\boldsymbol{\Phi}}_{j}^{11}$ 
($\widetilde{\boldsymbol{\Phi}}_{j}^{00}$) be the submatrix of 
$\boldsymbol{\Phi}_{-j,-j}$ with rows and columns corresponding to 
$\boldsymbol{\phi}_{-j,j}^1$ ($\boldsymbol{\phi}_{-j,j}^0$), so that we have 
\begin{align*}
    \boldsymbol{\Phi}_{-j,-j} = \begin{bmatrix} 
    \widetilde{\boldsymbol{\Phi}}_{j}^{11} & 
    \widetilde{\boldsymbol{\Phi}}_{j}^{10} \\ 
    \widetilde{\boldsymbol{\Phi}}_{j}^{01} & 
    \widetilde{\boldsymbol{\Phi}}_{j}^{00} \end{bmatrix},
\end{align*}
with $\widetilde{\boldsymbol{\Phi}}_{j}^{10} 
= (\widetilde{\boldsymbol{\Phi}}_{j}^{01})^\top$ after implicit row and 
column permutations. Finally, we define
\begin{align*}
    \xi_j = \phi_{jj} - \boldsymbol{\phi}_{-j,j}^\top 
    \boldsymbol{\Phi}_{-j,-j}^{-1} \boldsymbol{\phi}_{-j,j}= \phi_{jj}-
    (\boldsymbol{\phi}_{-j,j}^1)^\top \overline{\boldsymbol{\Phi}}_{j}^{-1} 
    \boldsymbol{\phi}_{-j,j}^1,
\end{align*}
where $\overline{\boldsymbol{\Phi}}_{j} = 
\widetilde{\boldsymbol{\Phi}}_{j}^{11} - 
\widetilde{\boldsymbol{\Phi}}_{j}^{10}
(\widetilde{\boldsymbol{\Phi}}_{j}^{00})^{-1}
\widetilde{\boldsymbol{\Phi}}_{j}^{01}$.

Following \citet{zhang2024incorporating}, we can show that $\xi_j$ and
$\boldsymbol{\phi}_{-j,j}^1$ are conditionally independent and 
\begin{align*}
    \xi_j \mid \boldsymbol{\Phi}_{-j,-j}, \boldsymbol{\alpha} 
    &\sim \mathrm{Gamma}\left(\frac{\eta(1+\zeta)+3}{2},\ 
    \frac{a_{\Phi,jj}}{2}\right), \\[6pt]
    \boldsymbol{\phi}_{-j,j}^1 \mid \boldsymbol{\Phi}_{-j,-j}, 
    \boldsymbol{\alpha} 
    &\sim \mathcal{N}\left(-\frac{\overline{\boldsymbol{\Phi}}_{j}
    \boldsymbol{a}_{\Phi,-j,j}^1}{a_{\Phi,jj}},\ 
    \frac{\overline{\boldsymbol{\Phi}}_{j}}{a_{\Phi,jj}}\right).
\end{align*}
The sample of $\phi_{jj}$ can be obtained
by $\phi_{jj} = \xi_j + (\boldsymbol{\phi}_{-j,j}^1)^\top 
\overline{\boldsymbol{\Phi}}_{j}^{-1} \boldsymbol{\phi}_{-j,j}^1$. To improve
computational efficiency, we adapt the accelerated sampling approach proposed
by \citet{zhang2024incorporating} for updating $\boldsymbol{\Phi}$.

\textit{(6) The conditional distributions of $\boldsymbol{x}_i$ for observations
in the training set}.
We use Gibbs sampling to sample $\boldsymbol{x}_i$ from its conditional 
distribution, which is
\begin{align*}
    \boldsymbol{x}_i \mid y_i, \boldsymbol{\beta}, \sigma^2, 
    \boldsymbol{\Gamma}, \boldsymbol{W}_i, \rho^2
    \sim \mathcal{N}\left(\boldsymbol{A}_{\boldsymbol{X}}
    \left(\frac{y_i\boldsymbol{\beta}}{\sigma^2} + 
    \frac{\boldsymbol{m}_{\boldsymbol{W}i}}{\rho^2}\right),\ 
    \boldsymbol{A}_{\boldsymbol{X}}\right),
\end{align*}
{\setlength{\emergencystretch}{3em}%
where $\boldsymbol{W}_i=\{\boldsymbol{w}_{ij}: j=1, \ldots, p\}$,
$\boldsymbol{m}_{\boldsymbol{W}i} = \left(\sum_{r=1}^{q_1}w_{i1r}, 
\ldots, \sum_{r=1}^{q_p}w_{ipr}\right)^\top$, 
$\boldsymbol{D}_{\boldsymbol{X}} = \mathrm{diag}(q_1/\rho^2, \ldots, 
q_p/\rho^2)$, and $\boldsymbol{A}_{\boldsymbol{X}} 
= (\boldsymbol{\beta}\boldsymbol{\beta}^\top/\sigma^2 + \boldsymbol{\Gamma} 
+ \boldsymbol{D}_{\boldsymbol{X}})^{-1}$.\par}

\textit{(7) The conditional distributions of $\boldsymbol{x}_i^*$ for
observations in the validation and test sets}. We use Gibbs sampling to sample
$\boldsymbol{x}_i^*$ from its conditional distribution, which is
\begin{align*}
    \boldsymbol{x}_i^* \mid \boldsymbol{\Gamma}, \boldsymbol{W}_i^*, \rho^2
    \sim \mathcal{N}\left(\boldsymbol{A}_{\boldsymbol{X}}
    ^*\frac{\boldsymbol{m}_{\boldsymbol{W}i}^*}{\rho^2},\ 
    \boldsymbol{A}_{\boldsymbol{X}}^*\right),
\end{align*}
where $\boldsymbol{W}_i^* = \{\boldsymbol{w}_{ij}^*:j=1, 
\ldots, p\}$ denotes the observed replicate covariate measurements for 
observation $i$ in the validation or test sets,
$\boldsymbol{m}_{\boldsymbol{W}i}^* = \left(\sum_{r=1}^{q_1}w_{i1r}^*,
\ldots, \sum_{r=1}^{q_p}w_{ipr}^*\right)^\top$, and 
$\boldsymbol{A}_{\boldsymbol{X}}^* = \left(\boldsymbol{\Gamma} + 
\boldsymbol{D}_{\boldsymbol{X}}\right)^{-1}$.

\textit{(8) The conditional distribution of $\rho^2$}. We use Gibbs sampling 
to sample $\rho^2$ from its conditional distribution, that is
\begin{align*}
    \rho^2 \mid \boldsymbol{X}, \boldsymbol{W} \sim \mathcal{IG}\left(a_\rho 
    + \frac{n\sum_{j=1}^p q_j}{2},\ b_\rho + 
    \frac{1}{2}\sum_{i=1}^n\sum_{j=1}^p\sum_{r=1}^{q_j}(w_{ijr} - 
    x_{ij})^2\right).
\end{align*}

\textit{(9) The conditional distribution of $\boldsymbol{\Gamma}$}. We use a 
block-wise Gibbs sampler to sample $\boldsymbol{\Gamma}$ column by column. Let
$\boldsymbol{S} = \boldsymbol{X}^\top \boldsymbol{X}$. For column $j$, we
partition $\boldsymbol{\Gamma}$, $\boldsymbol{S}$, and 
$\boldsymbol{\Upsilon}$ following the same partitioning scheme as in
\textit{(5)}, yielding, after implicit row and column permutations,
\begin{align*}
    \boldsymbol{\Gamma} = \begin{bmatrix} \boldsymbol{\Gamma}_{-j,-j} & 
    \boldsymbol{\gamma}_{-j,j} \\ \boldsymbol{\gamma}_{-j,j}^\top & 
    \gamma_{jj} \end{bmatrix}, \quad
    \boldsymbol{S} = \begin{bmatrix} \boldsymbol{S}_{-j,-j} & 
    \boldsymbol{s}_{-j,j} \\ \boldsymbol{s}_{-j,j}^\top & s_{jj} 
    \end{bmatrix}, \quad
    \boldsymbol{\Upsilon} = \begin{bmatrix} \boldsymbol{\Upsilon}_{-j,-j} & 
    \boldsymbol{\upsilon}_{-j,j} \\ \boldsymbol{\upsilon}_{-j,j}^\top & 0 
    \end{bmatrix}.
\end{align*} 
We define $\delta_j = \gamma_{jj} - \boldsymbol{\gamma}_{-j,j}^\top 
    \boldsymbol{\Gamma}_{-j,-j}^{-1} \boldsymbol{\gamma}_{-j,j}$.
Following \citet{wang2012bayesian}, we can show that $\delta_j$ and 
$ \boldsymbol{\gamma}_{-j,j}$ are conditionally independent and
\begin{align*}
    \delta_j \mid \boldsymbol{\Gamma}_{-j,-j}, \boldsymbol{X}, \iota_{jj}
    &\sim \mathrm{Gamma}\left(\frac{n}{2}+1,\ 
    \frac{s_{jj}+\iota_{jj}}{2}\right), \\ 
    \boldsymbol{\gamma}_{-j,j} \mid \boldsymbol{\Gamma}_{-j,-j}, 
    \boldsymbol{\Upsilon}, \boldsymbol{X}, \iota_{jj}
    &\sim \mathcal{N}\left(-\boldsymbol{C}_j\boldsymbol{s}_{-j,j},\ 
    \boldsymbol{C}_j\right),
\end{align*}
where $\boldsymbol{C}_j = \left[\boldsymbol{D}_{\boldsymbol{\Upsilon},j}^{-1} 
+ (s_{jj}+\iota_{jj})\boldsymbol{\Gamma}_{-j,-j}^{-1}\right]^{-1}$ with 
$\boldsymbol{D}_{\boldsymbol{\Upsilon},j} = \mathrm{diag}
(\boldsymbol{\upsilon}_{-j,j})$. We then recover $\gamma_{jj}$ as
$\gamma_{jj} = \delta_j + \boldsymbol{\gamma}_{-j,j}^\top 
\boldsymbol{\Gamma}_{-j,-j}^{-1} \boldsymbol{\gamma}_{-j,j}$.

\textit{(10) The conditional distributions of $\iota_{jk}$}. We use Gibbs 
sampling to sample $\iota_{jk}$ from its conditional distribution given 
$\boldsymbol{\Gamma}$, with $\upsilon_{jk}^2$ integrated out, which is
\begin{align*}
    \iota_{jk} \mid \gamma_{jk} &\sim \mathrm{Gamma}\left(a_\iota + 
    1,\ b_{\iota,jk} + |\gamma_{jk}|\right), \quad 1\leq j < k \leq p, \\
    \iota_{jj} \mid \gamma_{jj} &\sim \mathrm{Gamma}\left(a_\iota + 
    1,\ b_{\iota,jj} + \gamma_{jj}/2\right), \quad j=1,\dots, p.
\end{align*}

\textit{(11) The conditional distributions of $\upsilon_{jk}^2$}. We use Gibbs 
sampling to sample $\upsilon_{jk}^2$ from its conditional distribution, which 
is
\begin{align*}
    \frac{1}{\upsilon_{jk}^2} \mid \gamma_{jk}, \iota_{jk} \sim \mathcal{IN}
    \left(\frac{\iota_{jk}}{|\gamma_{jk}|},\ \iota_{jk}^2\right), \quad 1\leq j < k \leq p.
\end{align*}

\section[Generation of G0 and SigmaX]{Generation of $G_0$ and
$\boldsymbol{\Sigma}_X$}
\label{sec:appendixB}

$G_0$ is generated as follows. First, we generate $20$ pathways. Let genes
$1,\ldots,p_0$ be the important variables. The first pathway is composed of the
$p_0$ important variables and possibly a few unimportant ones, corresponding to
genes $1,\ldots,\max(N_1,p_0)$, where $N_1$ is drawn from a Poisson distribution
with mean $10$. The number of genes in other pathways are independently drawn
from the same Poisson distribution. The member genes of a pathway
are chosen randomly and independently of other pathways. Hence, the pathways can
be overlapping. Within each pathway, edges are generated according to the
following procedure. For each pair of genes $j$ and $k$ in the pathway:
\begin{enumerate}
    \item[(a)] If $1\le j,k\le p_0$ (both genes are important), generate an
    edge with probability $0.6$. Additionally, an edge is always included if
    $|j-k|=1$.
    \item[(b)] If one gene is important and the other is unimportant, do not
    generate an edge.
    \item[(c)] If both genes are unimportant, generate an edge with
    probability $0.3$.
\end{enumerate}
The adjacency matrix $G_0$ is constructed from the union of the edges generated
across all pathways.

Given $G_0$, the covariance matrix $\boldsymbol{\Sigma}_X$ is designed as
follows.
\begin{enumerate}
    \item[(i)] Initialize $\boldsymbol{\Omega}=\boldsymbol{I}_p$.
    \item[(ii)] Calculate the vertex degrees $D_j=\sum_{k=1}^{p}G_{0,jk}$.
    \item[(iii)] For each pair $j<k$ with $G_{0,jk}=1$, set 
    $\Omega_{jk}=\Omega_{kj}=-\theta_{jk}/\{\max(D_j,D_k)\times 1.1+0.1\}$
    where $\theta_{jk}=1$ if $1\le j,k\le p_0$, and otherwise $\theta_{jk}$ is
    independently drawn from $\{-1,1\}$ with equal probability.
    \item[(iv)] Set $\boldsymbol{\Sigma}_X=\boldsymbol{\Omega}^{-1}$ and
    rescale it so that all diagonal elements equal $1$.
\end{enumerate}
Note that $\boldsymbol{\Omega}$ is symmetric and strictly diagonally dominant,
so the resulting covariance matrix $\boldsymbol{\Sigma}_X$ is positive definite.
Moreover, $x_{ij}$ and $x_{ik}$ are partially correlated only if $G_{0,jk}=1$,
for any observation $i$.

\section*{Acknowledgements}

This work was supported by the National Institutes of Health, RF1 AG063481 and
R01 AG071174. The content is solely the responsibility of the authors and does
not necessarily represent the official views of the National Institutes of
Health.

{\setlength{\emergencystretch}{3em}%
The complete ADNI Acknowledgment is available at
\url{https://adni.loni.usc.edu/wp-content/uploads/how_to_apply/ADNI_Acknowledgement_List.pdf}
\par}

\vspace*{-8pt}

\section*{Data Availability}
The data that support the findings in this paper are available in the ADNI 
database at \url{https://adni.loni.usc.edu} and in the Gene Expression 
Omnibus under accession number GSE24080 
at \url{https://www.ncbi.nlm.nih.gov/geo/query/acc.cgi?acc=GSE24080}.
The \pkg{badgers} R package and source code are available at
\url{https://github.com/zyhu888/badgers} \citep{badgers2026}.

\singlespacing
\bibliographystyle{chicago}
\bibliography{reference}

\end{document}